\documentclass[journal,comsoc]{IEEEtran}
\IEEEoverridecommandlockouts
\usepackage{cite}
\usepackage{bm}
\usepackage{amsmath,amssymb,amsfonts}
\usepackage{graphicx}
\usepackage{multirow}
\usepackage{bm}
\usepackage{float}
\usepackage{amsmath,amsfonts}
\usepackage{algorithmic}
\usepackage{algorithm}
\usepackage{array}
\usepackage[caption=false,font=normalsize,labelfont=sf,textfont=sf]{subfig}
\usepackage{textcomp}
\usepackage{stfloats}
\usepackage{url}
\newtheorem{lemma}{Lemma}[section]
\newenvironment{proof}{ \paragraph*{\hspace{-1em}Proof}}{\hfill$\square$}
\usepackage{verbatim}
\usepackage{cite}
\usepackage{amsmath, cite}
\usepackage{graphicx}
\usepackage{subcaption}
\usepackage{enumitem}

\usepackage{array}
\usepackage{textcomp}
\usepackage{epstopdf}
\usepackage[font=small,labelfont=bf,justification=centering]{caption}
\usepackage[table]{xcolor}
\usepackage{booktabs}
\usepackage{subcaption}
\usepackage{mathtools}
\DeclarePairedDelimiter\ceil{\lceil}{\rceil}

\usepackage[printonlyused,withpage]{acronym}
\acrodef{BS}{base station}
\acrodef{BB}{base-band}
\acrodef{CSI}{channel state information}
\acrodef{ZF}{zero-forcing}
\acrodef{UE}{user equipment}
\acrodef{UL}{uplink}
\acrodef{DL}{downlink}
\acrodef{TDD}{time-division duplexing}
\acrodef{FDD}{frequency-division duplexing}
\acrodef{ZF}{zero-forzing}
\acrodef{LMMSE}{linear minimum mean squared error}
\acrodef{MRC}{maximum ratio combining}
\acrodef{BBU}{baseband unit}
\acrodef{DPD}{digital pre-distortion}
\acrodef{OTA}{over-the-air}
\acrodef{SNR}{signal to noise ratio}
\acrodef{TX}{transmit}
\acrodef{RX}{receive}
\acrodef{AWGN}{additive white Gaussian noise}
\acrodef{MIMO}{multiple-input multiple-output}
\acrodef{LS}{least-squares}
\acrodef{MVU}{minimum-variance unbiased}
\acrodef{EM}{expectation-maximization}
\acrodef{MSE}{mean square error}
\acrodef{CRLB}{Cramer-Rao lower bound}
\acrodef{LIS}{large intelligent surface}
\acrodef{LOS}{line of sight}
\acrodef{MLP}{memory-less polynomial}
\acrodef{SNDR}{signal-to-noise-plus-distortion-ratio}

\begin{document}
\title{Hardware Distortion Modeling for Panel Selection in Large Intelligent Surfaces  \\
}
\author{\IEEEauthorblockN{Ashkan Sheikhi,~\IEEEmembership{Member,~IEEE,} Juan Vidal Alegría,~\IEEEmembership{Member,~IEEE,} and Ove Edfors,~\IEEEmembership{Senior Member,~IEEE}.}
\thanks{This work was supported by "SSF Large Intelligent Surfaces - Architecture and Hardware" Project CHI19-0001.}
\IEEEauthorblockA{\textit{Department of Electrical and Information Technology, Lund University, Lund, Sweden.} \\
		Email: \{ashkan.sheikhi, juan.vidal\_alegria, ove.edfors\}@eit.lth.se}
}
\maketitle

\begin{abstract}
Hardware distortion in large intelligent surfaces (LISs) may limit their performance when scaling up such systems. It is of great importance to model the non-ideal effects in their transceivers to study the hardware distortions that can affect their performance. Therefore, we have focused on modeling and studying the effects of nonlinear RX-chains in LISs. We first derive expressions for SNDR of a LIS with a memory-less polynomial-based model at its RX-chains. Then we propose a simplified double-parameter exponential model for the distortion power and show that compared to the polynomial based model, the exponential model can improve the analytical tractability for SNDR optimization problems. In particular, we consider a panel selection optimization problems in a panel-based LIS scenario and show that the proposed model enables us to derive two closed-form sub-optimal solutions for panel selection, and can be a favorable alternative to high-order polynomial models in terms of computation complexity, especially for theoretical works on hardware distortion in \ac{MIMO} and LIS systems. Numerical results show that the sub-optimal closed-form solutions have a near-optimal performance in terms of SNDR compared to the global optimum found by high-complexity heuristic search methods.
\end{abstract}

\begin{IEEEkeywords}
Hardware Distortion, Large Intelligent Surface, MIMO, Panel Selection.
\end{IEEEkeywords} 

\section{Introduction}
The rapid increase in number of devices and quality of service demands within 5G and forthcoming 6G networks has led to a substantial escalation in overall network requirements. To address these demands, the development and implementation of novel physical layer technologies are essential for accommodating the heightened performance expectations and ensuring network reliability and efficiency. \Acp{LIS} are regarded as a pivotal advancement in the ongoing evolution of wireless communication networks.  In theory, \acp{LIS} have demonstrated significant potential to meet these expectations, primarily by offering greater degrees of freedom compared to conventional massive MIMO systems \cite{HuSha2017,ShaHuBeyond1,Dardari}. While the anticipated benefits of deploying \ac{LIS} technology are promising, considerable debate persists regarding the feasibility and hardware implementation challenges associated with integrating \ac{LIS} into future wireless networks. \acp{LIS} are envisioned to have hundreds to thousands of active transceiver chains which can result in a huge leap in the implementation cost, power consumption, and overall processing complexity of the system \cite{TWC2024}.

Achieving the expected theoretical gains when implementing \acp{LIS} in future wireless networks may only become feasible by deploying less expensive hardware components in the transceiver chains. The drawback from selecting these low-cost components are mainly the non-ideal effects that can introduce hardware distortion in the system, which can degrade the overall performance \cite{Juan2019}. The inter-play between system performance, non-linearity, power consumption, and hardware complexity is one of the most important aspects in the design of future wireless transceivers \cite{AshkanICC2021,Muris2021}. Mitigating non-ideal hardware distortion is expected to be an important subject when implementing \acp{LIS} for future wireless networks.

In addition, to save on resources, we would also like to activate as few transceiver chains as possible. Full control of the activation of individual transceivers means full flexibility, but also lead to excessive system complexity. To limit complexity, we can arrange antennas in panels where all transceiver chains in a panel are switched on or off at the same time, which we call panel-based \ac{LIS}. While there are many classic models and methods to model, study, and compensate the hardware distortion effects in MIMO transceivers \cite{Morgan,Tehrani2010,WCL2024,AshkanWCNC2023}, applying them to LIS scenarios generally results in high-complexity problems when designing and optimizing the system. For example, while polynomial models are of great interest in the MIMO literature, employing them in a panel-selection optimization for a panel-based LIS, results in high complexity problems which can only be solved by heuristic methods \cite{TWC2024}. Therefore, there is a need for more analytically favorable models for hardware distortion in \acp{LIS} with negligible loss in the performance.

In this paper, we study the problem of receiver hardware distortion in \acp{LIS} with non-linear RX-chains. We first analyze the distortion effect for the memory-less polynomial (MLP) model and derive the SNDR for \ac{MRC} scheme. Then, we propose a double-parameter exponential model for the hardware distortion power to reduce the complexity of the SNDR optimization problems. In particular, we formulate the problem of panel selection in panel-based \acp{LIS} and show that with the MLP model, the panel selection problem can only be solved by heuristic search. On the other hand, the proposed model leads to more tractable optimization problems. We will show that the simplified panel selection problem can be approximated and solved into close-form sub-optimal solutions which can be adopted for the original panel selection problem, with near-optimum performance in terms of the \ac{SNDR}.
 
\section{System Model} \label{SystemModel}
We consider an uplink scenario where a single-antenna \ac{UE} is served by a \ac{LIS} through a narrow-band \ac{LOS} channel\footnote{We have considered a single-user case to isolate the effect of hardware distortion at the \ac{LIS} RX-chains from other non-ideal effects such as inter-user interference, since hardware distortion is the main focus of this work.}. The \ac{LIS} consists of $N\gg 1$ antenna elements with non-linear RX-chains. The $N\times 1$ received vector at the LIS is
\begin{align}\label{rSystenModel}
    \bm{r}=f\left(\bm{h}s\right)+\bm{n},    
\end{align}
where $s \in \mathbb{C}$, with $\mathbb{E}\{|s|^2\}=P$, is the \ac{BB} symbol transmitted by the \ac{UE}, and $\bm{h} \in \mathbb{C}^{N\times1}$ is the \ac{LOS} channel vector. The component-wise function $f(\cdot): \mathbb{C}^{N\times1} \rightarrow \mathbb{C}^{N\times1}$ models the overall hardware distortion effects of the \ac{LIS} non-linear RX-chains, and $\bm{n}\sim\mathcal{CN}(\bm{0},\sigma^2\bm{I}_N)$ models the receiver thermal noise.

\subsection{RX-chain non-linearity}
To analyze the effect of RX-chain hardware distortion, we can transform the RX-chain output $\bm{z} \triangleq f\left(\bm{h}s\right)$ into an additive form by leveraging the \ac{LMMSE} of $\bm{z}$ given $\bm{x}\triangleq\bm{h}s$, which is
\begin{align} \label{LMMSE}
    \bm{z}=\bm{C}_{\bm{zx}}\bm{C}_{\bm{xx}}^{-1}\bm{x}+\bm{\eta},
\end{align}
where $\bm{\eta}$ is the estimation error. This is the same technique as applying the Bussgang theorem to the non-linearity function \cite{DemirBussgang}. We can therefore re-write $\bm{r}$ as
\begin{align} \label{Bussgang}
    \bm{r}=\bm{G}\bm{h}s+\bm{\eta}+\bm{n},
\end{align}
where $\bm{G}=\bm{C}_{\bm{zx}}\bm{C}_{\bm{xx}}^{-1}$. If we assume that the output of each RX-chain depends solely on its input and it is independent of other RX-chains, i.e., $\bm{z}_n$ only depends on $\bm{x}_n$, we have $\bm{G}=\text{diag}\left\{g_n\right\}$, $\bm{C_{\eta\eta}}=\text{diag}\left\{C_n\right\}$, with $g_n$ and $C_n$ corresponding to the Busssgang gain compression and distortion power for the $n$'th antenna. 

One of the most widely used models for non-linearities in wireless transceivers is the memory-less polynomial model \cite{Morgan,Razavi2011}, given by
\begin{align}\label{memoryLessModel}
	f(x_n) = \sum_{k=0}^{L-1} a_{2k+1}~x_n|x_n|^{2k},
\end{align}
where $x_n$ is the input to one of the RX-chains of the \ac{LIS}, and $a_{2k+1}$ are the model parameters. The model coefficients can be calculated by curve fitting to input-output measurements data from transceivers \cite{Ericsson2016} for a limited range of input amplitude. For a Gaussian input $x_n\sim\mathcal{CN}(0,\rho_n)$ which is a high peak-to-average signal, we need to normalize the model coefficients and consider a sufficient back-off at the RX-chains. The Bussgang parameters $\forall n \in\{1,\dots,N\}$ can then be calculated as \cite{TWC2024}
\begin{align}
    &g_n =\sum_{k=0}^{L-1}a_{2k+1}(k+1)!\rho_n^k, \label{BussgangGC} \\
    &C_{n}=\sum_{k=1}^{2L-1}\left(k!\rho_n^k\sum_{i=1}^{k}a_{2i-1}\bar{a}_{2k-2i+1}\right)-|g_n|^2\rho_n, \label{BussgangCetaeta}
\end{align}
which we will use to analyze the \ac{SNDR} in the reminder of this paper.

\subsection{SNDR for \ac{MRC}}
Let us assume that the \ac{LIS} employs a combining vector $\bm{v}$ to equalize the received signal $\bm{r}$. It has been shown that \ac{MRC} can leverage the available spatial degrees of freedom\cite{ShaHuBeyond1} effectively in \ac{LIS} scenarios, and it is more favorable due to its reduced complexity.
In our system model, we have an effective channel given by $\tilde{\bm{h}} = \bm{G}\bm{h}$. Therefore, the \ac{MRC} vector is expressed as $\bm{v}^T = \tilde{\bm{h}}^H / \Vert\tilde{\bm{h}}\Vert$. This effective channel accounts for both the physical channel and the multiplicative hardware distortion effects. Since the signals used for channel estimation are also influenced by hardware distortion, the \ac{UL} \ac{UE} pilots would only allow the LIS to estimate the effective channel $\tilde{\bm{h}}$ \cite{TWC2024}. For the purposes of this analysis, we assume that the \ac{LIS} has a perfect estimate of $\tilde{\bm{h}}$.

By applying the \ac{MRC} combining vector $\bm{v}^T=\tilde{\bm{h}}^H/\|\tilde{\bm{h}}\|$ to the received signal $\bm{r}$, while taking into account the Bussgang decomposition from \eqref{Bussgang}, we can calculate the \ac{SNDR} as
\begin{align}\label{SNDRMISO}
    \gamma = \frac{P\sum_{n=1}^{N}\left|\tilde{h}_n\right|^2}{\frac{\sum_{n=1}^{N}C_n\left|\tilde{h}_n\right|^2}{\sum_{n=1}^{N}\left|\tilde{h}_n\right|^2}+\sigma^2}.
\end{align}
For the memory-less polynomial model we have 
\begin{align}
    &\left|\tilde{h}_n\right|^2=\left|{h}_n\right|^2\left|\sum_{k=0}^{L-1}a_{2k+1,n}(k+1)!~P^k|h_n|^{2k}\right|^2, \\
    &C_n=\sum_{k=1}^{2L-1}\left(k!~P^k|h_n|^{2k}\sum_{i=1}^{k}a_{2i-1,n}\bar{a}_{2k-2i+1,n}\right) - \left|\tilde{h}_n\right|^2P,
\end{align}
which are calculated according to \eqref{BussgangGC} and \eqref{BussgangCetaeta}.

\subsection{Exponential Model for Distortion Power}
While it is possible to derive closed-form expressions for \ac{SNDR} with the memory-less polynomial model \cite{TWC2024}, a simplified model with fewer number of parameters is of fundamental interest, especially when dealing with optimization problems involving the SNDR. The model should be an increasing function of the input power, and follow the general form of the distortion power \eqref{BussgangCetaeta}. Another downside with the distortion power function for memory-less polynomial model is that the distortion power may increase unboundedly when input power grows, an effect that can not occur in reality. In particular, when considering high peak-to-average power signals, such as Gaussian symbols, this can have a significant influence on the accuracy of the analysis. Therefore, the proposed model for the distortion power should not grow unboundedly with the input power.

With the given conditions on the model properties, we propose the following double-parameter exponential model for the distortion power
\begin{align}\label{expModel}
    \Tilde{C}_{n} = \rho \left(1-e^{-\beta \rho^q}\right),
\end{align}
where $\beta>0$ and $q>2$ are the model parameters and are calculated by curve-fitting to hardware measurements, or potentially, to the distortion power of any other well-known model such as \eqref{BussgangCetaeta} for memory-less polynomial model. We will show that this model can be exploited as a tool to simplify the \ac{SNDR} optimization problems for systems with hardware distortion. In particular, we will show an application of this model in the \ac{LIS} panel selection problem in the next section. 

\section{Panel Selection in LIS} \label{Sec_Panel_Selection}
\begin{figure}[t]
	\centering
	\includegraphics[width=3.4in]      {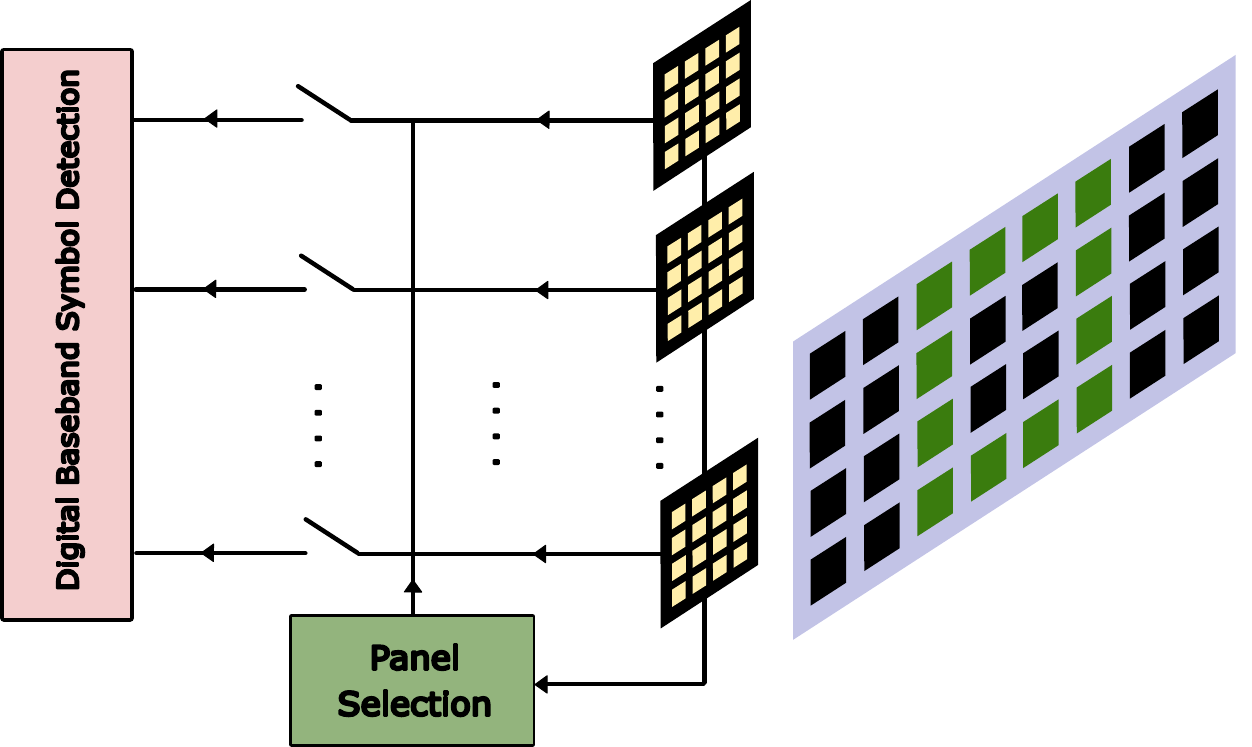}
	\caption{LIS configuration and Panel Selection. Each panel, represented by a square, has the same number of antennas and green squares indicate the active panels. Each antenna is equipped with a non-linear analogue front end (AFE).}
	\label{scheme}
\end{figure}
One of favorable approaches in terms of practicality for \ac{LIS} deployment is to construct them as a grid of panels \cite{Andreia}. We consider a panel-based \ac{LIS} with $N_p$ panels each consisting of $M$ antenna elements with non-linear RX-chains. We assume that there is a resource constraint in the system which forces the LIS to only use $N_\text{max}$ panels for received signal combining. Fig. \ref{scheme} illustrates an example system architecture with panel selection. We select the \ac{SNDR} after \ac{MRC} as the objective function and formulate a panel selection problem with the aim of finding the set of $N_\text{max}$ panels to achieve the highest \ac{SNDR}.

The panel selection problem can be translated into the following optimization problem
\begin{align} \label{PanelOptProb}
	\max_{z_n} ~ &\frac{PM\sum_{n=1}^{N_p}z_n\left|\tilde{h}_n\right|^2}{\frac{\sum_{n=1}^{N_p}z_nC_n\left|\tilde{h}_n\right|^2}{\sum_{n=1}^{N_p}\left|\tilde{h}_n\right|^2}+\sigma^2},\\
	&\mathrm{s.t.}~~ z_n\in \{0,1\} ~~~~~~~ n=1,2,...,N\nonumber \\
	&~~~~~ \sum_{n=1}^{N_p}z_n\leq N_\text{max}, \nonumber
\end{align}
where $z_n$ is the binary variable for panel selection, and $|\tilde{h}_n|^2$ is the effective channel gain between the antenna elements on the $n$'th panel and the UE. The considered formulation of the problem assumes that the distance between antenna elements of each panel is negligible compared to the \ac{UE}-\ac{LIS} distance, which corresponds to having the users in the far-field of each LIS-panel, while they can still be in the near-field of the whole LIS. In general, we are able to solve this problem by heuristic search methods for any hardware distortion model.

An alternative approach, equivalent to a large extent to the original problem, is to consider a SISO case and find the optimum received power for maximum SNDR, and then select the $N_\text{max}$ panels with the closest received power to that optimum value. For the memory-less polynomial model, the panel selection problem is reduced to
\begin{align} \label{SISOPolytOpt}
	\max_{\rho} ~ &\frac{|\sum_{k=0}^{L-1}a_{2k+1}(k+1)!\rho^k|^2\rho}{\sum_{k=1}^{2L-1}\left(k!\rho^k\sum_{i=1}^{k}a_{2i-1}\bar{a}_{2k-2i+1}\right)-|g|^2\rho+\sigma^2}\\
	&\mathrm{s.t.}~~  0<\rho<\rho_{\max}, \nonumber&
\end{align}
which is still not analytically tractable. We are thus interested in approximating the objective function in this optimization problem to find effective closed-form solutions.

\begin{figure}[t]
	\centering
	\includegraphics[width=3.3in]{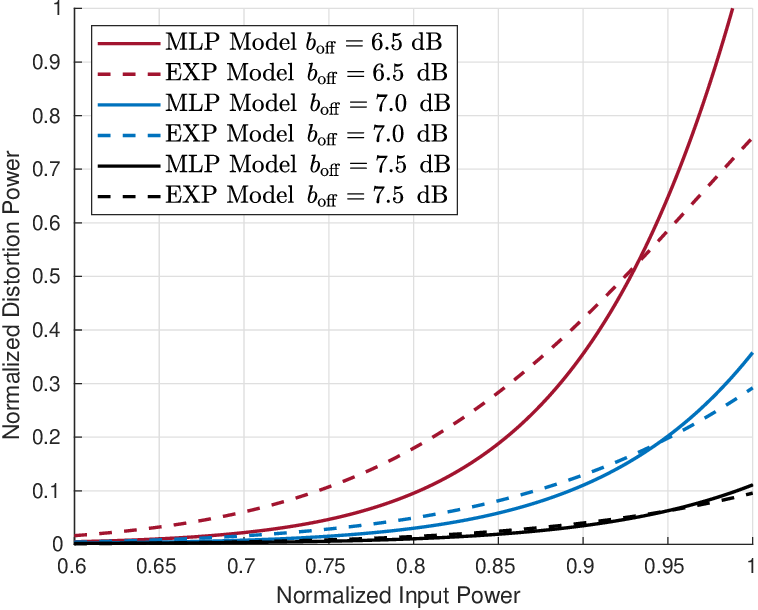}
	\caption{MLP and EXP Distortion Models.}
	\label{EXPMLP}
\end{figure}
The panel selection problem defined above constitutes a good application of the proposed distortion model, which may help increasing tractability. We approximate the objective function in \eqref{SISOPolytOpt} by deploying the proposed exponential model in \eqref{expModel}. We also neglect the gain compression parameter in the numerator since it has a negligible effect on \ac{SNDR} compared to the distortion power \footnote{We have studied and verified this numerically for the measurement data from \cite{Ericsson2016}. We will also show that this assumption has negligible impact on the performance of the proposed sub-optimal solutions.}. After applying these assumptions, we end-up with the following simplified problem.
\begin{align} \label{SISOOpt}
	\max_{\rho} ~ &\frac{\rho}{\rho \left(1-e^{-\beta \rho^q}\right)+\sigma^2}\\
	&\mathrm{s.t.}~~  0<\rho<\rho_{\max}, \nonumber&
\end{align}
We will show that, unlike \eqref{SISOPolytOpt}, this problem can be solved into closed-form approximated solutions with near-optimal performance.
\begin{lemma} \label{ExpApprox12}
The closed-form solution to the optimization problem \eqref{SISOOpt} can be approximated by either of the following optimal values for $\rho_{\text{opt}}$ with negligible error.
\begin{align}
    &\rho_{\text{opt}_1} \approx \left({\frac{1-\sqrt{1-\frac{4\sigma^2}{q}}}{2\beta}}\right)^{1/(q+1)},  \\
    &\rho_{\text{opt}_2} \approx \text{exp}\left({\frac{\text{Ln}(\sigma^2)-\text{Ln}(q\beta)}{q+1}}\right).
\end{align}
\begin{proof}
See Appendix \ref{ProofsAppendix}.
\end{proof}
\end{lemma}

We will study the performance of these approximated sub-optimal solutions and show that both have negligible error compared to the global optimum, while $\rho_{\text{opt}_2}$ is generally more accurate. The LIS can thus exploit these closed-form results for efficient panel selection. Although this is a sub-optimal approach to deal with the original problem in \eqref{PanelOptProb}, we will show that the performance is very close to the global optimal case achieved with heuristic search methods, which means that we can get a near-optimum solution with significantly lower computation complexity by adopting the proposed distortion model and approximated panel selection solutions.

\section{Numerical Results}
\begin{figure}[t]
	\centering
	\includegraphics[width=3.3in]{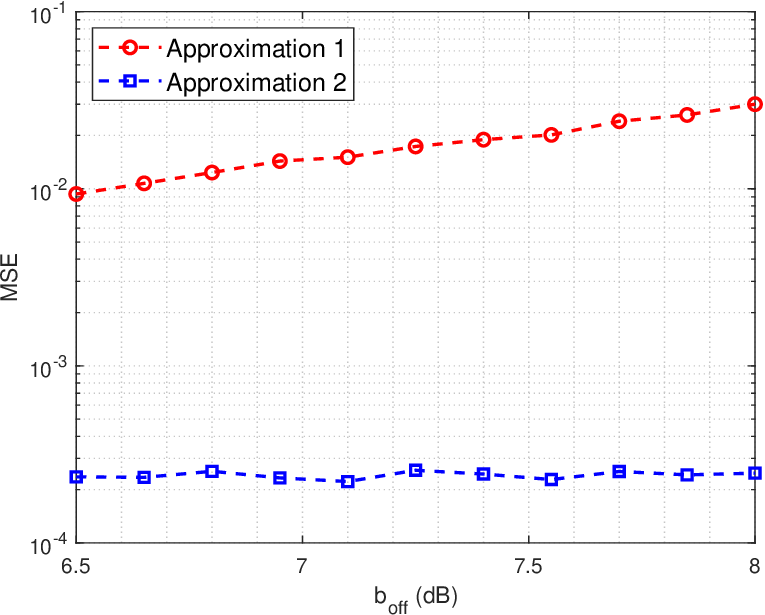}
	\caption{Mean square error (MSE) for optimum $\rho$ approximations from lemma \ref{ExpApprox12}.} 
	\label{MSE}
\end{figure}
In this section, we analyze the hardware distortion in \acp{LIS} for the memory-less polynomial model (MLP) and the proposed exponential model (EXP). For the RX-chains non-linearity model parameters, we have considered an 11-th order memory-less polynomial model based on \cite{Ericsson2016} for a Gallium Nitride (GaN) amplifier operating at $2.1$ GHz at a sample rate of 200 MHz and a signal bandwidth of 40 MHz. Different levels of back-off are considered which adjust the severeness of the distortion as described in \cite{TWC2024}.

In Fig. \ref{EXPMLP} we have compared the approximated EXP distortion power \eqref{expModel} to the exact MLP distortion power \eqref{BussgangCetaeta} for the GaN amplifier with three different levels of back-off. The EXP model parameters $\beta$ and $q$ calculated with MATLAB curve fitting. We can see that the approximated EXP model follows the exact model closely, and the difference is lower for higher values of back-off. We will show that this difference has negligible effect on the performance of the proposed sub-optimal solutions for panel selection.

In Fig. \ref{MSE} we illustrate the \ac{MSE} of the proposed approximations in Lemma \ref{ExpApprox12} with respect to the accurate optimum solution to problem \eqref{SISOOpt}, found by heuristic search. We can see that both approximations have very low MSE, while the second approximation performs better for all levels of back-off. Therefore, we focus on the second approximation for solving the approximate version of the panel selection problem \eqref{PanelOptProb}.

\begin{figure}[t]
	\centering
	\includegraphics[width=3.3in]{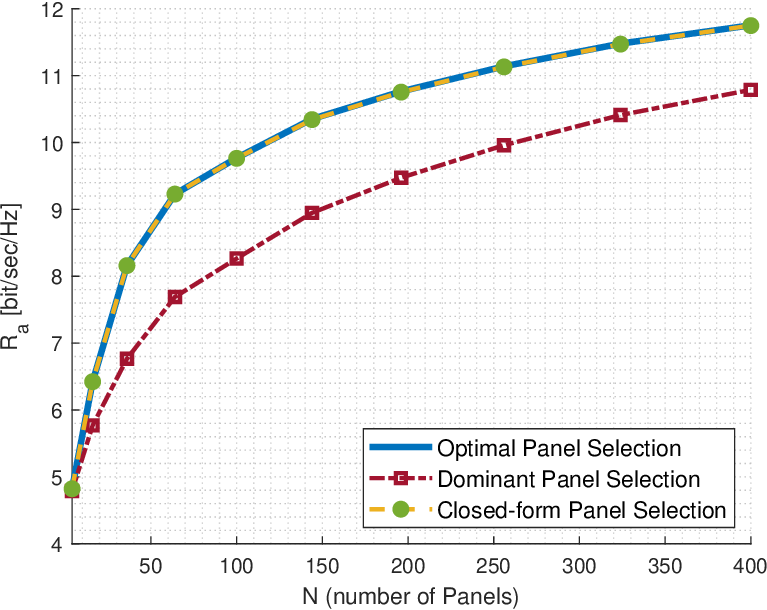}
	\caption{SE lower bound vs number of panels. The UE is at distance $d=50\lambda$ from the center of LIS transmitting with power $P$ such that SNR$=10dB$ at the center of LIS. Each panel is equipped with $M=16$ antenna elements with $\lambda/2$ spacing, $N_\text{max} = \ceil{0.1N}$, $b_\text{off}=7dB$, and there is a distance of $\delta_p=5\lambda$ between the center of adjacent panels.} 
	\label{PS_CL_SE_vs_R}
\end{figure}
Fig. \ref{PS_CL_SE_vs_R} illustrates the performance of LIS, in terms of lower bound on spectral efficiency (SE), when adopting the proposed sub-optimal solution for panel selection compared to the global optimum case found by solving \eqref{PanelOptProb} with heuristic search. We have also included a base-line approach where the \ac{LIS} only selects the panels with highest channel gain which is the optimal solution in ideal systems. Firstly, we can see that in the presence of RX-chain hardware distortion, panel selection can improve the system performance significantly. Secondly, we can see that the proposed sub-optimal solution based on the second approximated power level from Lemma \eqref{ExpApprox12} has near-perfect performance.

\section{Conclusion}
In this paper, we have studied the hardware distortion effects in \acp{LIS} with non-ideal RX-chains. Firstly, we considered the memory-less polynomial model and formulated the SNDR for the LIS. Although the SNDR can be characterized in closed-form this model, the resulting expression gives poor analytical tractability towards solving optimization problems such as \ac{LIS} panel selection. Thus, we have proposed a double-parameter exponential model for the distortion power and employed it to characterize close-form approximate solutions to the \ac{LIS} panels. The proposed methods attain close-to-optimum performance, essentially overlapping the performance achieved by global optimum solutions found through heuristic search.

{\appendices
\section{Proof of Approximations}\label{ProofsAppendix}
To solve \eqref{SISOOpt}, we can show that the problem is convex\footnote{The proof is not included but it is straightforward.}. We can therefore solve it by finding the roots of the first derivative of the objective function. The equation to find the roots can be simplified to
\begin{align} \label{derivative}
    \sigma^2-q\beta\rho^{q+1}e^{-\beta \rho^q} = 0.
\end{align}
By defining $x\triangleq\beta\rho^{q+1}$ and approximating $e^{-\beta \rho^q}$ with $e^{-x}$, we have
\begin{align} \label{eqApprox}
    \sigma^2-qxe^{-x} \approx 0.
\end{align}
Fitting the proposed exponential model to the measurement data from \cite{Ericsson2016} shows that in all cases $\beta \ll1$, which results in $x\ll1$. Therefore, we can use the Taylor approximation to write $qxe^{-x}\approx x(1-x)$. Replacing this in \eqref{eqApprox} results in a quadratic equation which gives the first approximated solution $\rho_{\text{opt}_1}$.

The second approximation $\rho_{\text{opt}_2}$ can be found by taking the natural logarithm of \eqref{derivative} which results in
\begin{align} \label{Lnderivative}
    \text{Ln}(\sigma^2) = (q+1)\text{Ln}(\rho)+\text{Ln}(q\beta)-\beta\rho^q.
\end{align}
As discussed above, $\beta\rho^q\ll1$, therefore we can neglect the last term, and solve the above equation for $\rho$ which results in the closed form approximated solution $\rho_{\text{opt}_2}$.
}
\bibliographystyle{IEEEtran}
\bibliography{IEEEabrv,Refs}
\end{document}